\documentclass[conference]{IEEEtran}
\IEEEoverridecommandlockouts
\usepackage{cite}
\usepackage{amsmath,amssymb,amsfonts}
\usepackage{algorithmic}
\usepackage{graphicx}
\usepackage{textcomp}
\usepackage{xcolor}
\usepackage{hyperref}
\usepackage{booktabs} 
\usepackage{tabularx} 
\usepackage{caption} 
\usepackage{subcaption} 
\usepackage{pgfplots} 
\pgfplotsset{compat=1.17} 

\def\BibTeX{{\rm B\kern-.05em{\sc i\kern-.025em b}\kern-.08em
    T\kern-.1667em\lower.7ex\hbox{E}\kern-.125emX}}

\begin{document}

\title{MetricSynth: Framework for Aggregating DORA and KPI Metrics Across Multi-Platform Engineering}

\author{\IEEEauthorblockN{Pallav Jain, Yuvraj Agrawal, Ashutosh Nigam, Pushpak Patil}
\IEEEauthorblockA{\textit{Adobe Inc.}}
}

\maketitle

\begin{abstract}
In modern, large-scale software development, engineering leaders face the significant challenge of gaining a holistic and data-driven view of team performance and system health. Data is often siloed across numerous disparate tools, making manual report generation time-consuming and prone to inconsistencies. This paper presents the architecture and implementation of a centralized framework designed to provide near-real-time visibility into developer experience (DevEx) and Key Performance Indicator (KPI) metrics for a software ecosystem. By aggregating data from various internal tools and platforms, the system computes and visualizes metrics across key areas such as Developer Productivity, Quality, and Operational Efficiency. The architecture features a cron-based data ingestion layer, a dual-schema data storage approach, a processing engine for metric pre-computation, a proactive alerting system, and utilizes the open-source BI tool Metabase for visualization, all secured with role-based access control (RBAC). The implementation resulted in a significant reduction in manual reporting efforts, saving an estimated 20 person-hours per week, and enabled faster, data-driven bottleneck identification. Finally, we evaluate the system's scalability and discuss its trade-offs, positioning it as a valuable contribution to engineering intelligence platforms.
\end{abstract}


\section{Introduction}
The complexity of managing software engineering teams for a software product which spans multiple platforms, presents a formidable challenge. Engineering leaders and stakeholders require timely and accurate insights to make informed, data-driven decisions. However, the necessary data is typically fragmented across a multitude of specialized tools, including source control systems, project management platforms, continuous integration services, log analysis tools, and application performance monitoring solutions. This fragmentation necessitates significant manual effort to collect, aggregate, and analyze data, a process that is not only inefficient but also susceptible to error and inconsistency.

Prior to the implementation of our unified framework, the process of generating performance reports was a significant operational burden. Engineering and program managers would spend hours manually sourcing data from disparate systems and creating complex pivot tables. This manual process lacked data interconnectedness, making it nearly impossible to draw correlations between, for instance, a spike in CI/CD build failures and a subsequent slowdown in feature delivery.

To address these challenges, we have developed a centralized framework that provides a unified, near-real-time view into developer experience (DevEx) and key performance indicators (KPIs). Our work is heavily influenced by industry standards such as the DevOps Research and Assessment (DORA) metrics, which are strong indicators of high-performing teams \cite{ b7, b_forsgren_2018}. We track these alongside a curated set of internal KPIs tailored to our organizational goals across five key areas: Developer Productivity, Quality \& Reliability, Automation Tooling, Engagement, and Operational Efficiency.

The main contributions of this paper are:
\begin{itemize}
    \item The design of a scalable, multi-layered architecture for aggregating engineering data, featuring a dual-schema storage strategy and a pre-computation layer for performance.
    \item A curated framework of DORA-aligned and internal KPI metrics tailored for a large-scale, multi-platform software product.
    \item An empirical evaluation of the framework's impact, including quantitative time savings and qualitative case studies on improving development workflows.
    \item A discussion of the system's unique aspects, including its extensibility via a secure API and its foundation on an open-source stack, offering a blueprint for similar initiatives.
\end{itemize}

\section{Related Work}
The field of engineering intelligence has seen significant growth, with various tools and research aiming to provide visibility into the software development lifecycle. Our work is positioned within this landscape.

\textbf{Commercial Engineering Intelligence Platforms:} Tools like Hatica, LinearB, and Jellyfish \cite{b21} offer sophisticated platforms by integrating with common development tools. While powerful, they often come with significant licensing costs and may offer limited customizability for organization-specific metrics or data sources. Our approach, utilizing an open-source core (Metabase, MongoDB), provides a high degree of flexibility and extensibility at a lower operational cost.

\textbf{Platform-Integrated Analytics:} Tools like GitLab \cite{b11} and GitHub offer built-in analytics, including DORA metrics. These are valuable but are often confined to the data within their own ecosystem. Our framework's primary contribution is its ability to unify data from a heterogeneous, multi-vendor toolchain (Jira, Jenkins, Splunk, Firebase, etc.), providing a holistic view that platform-specific tools cannot.

\textbf{Academic Research:} Research in software analytics has long focused on mining software repositories to understand developer productivity and software quality. Our system builds on these principles, applying them in a large-scale industrial context. Our unique contribution is the detailed architectural blueprint for a multi-platform system that balances industry-standard metrics (DORA) with deeply contextual internal KPIs, and a discussion of practical challenges like scalability and Goodhart's Law.

\section{System Architecture}
\begin{figure*}[htbp]
\centering
\includegraphics[width=\textwidth]{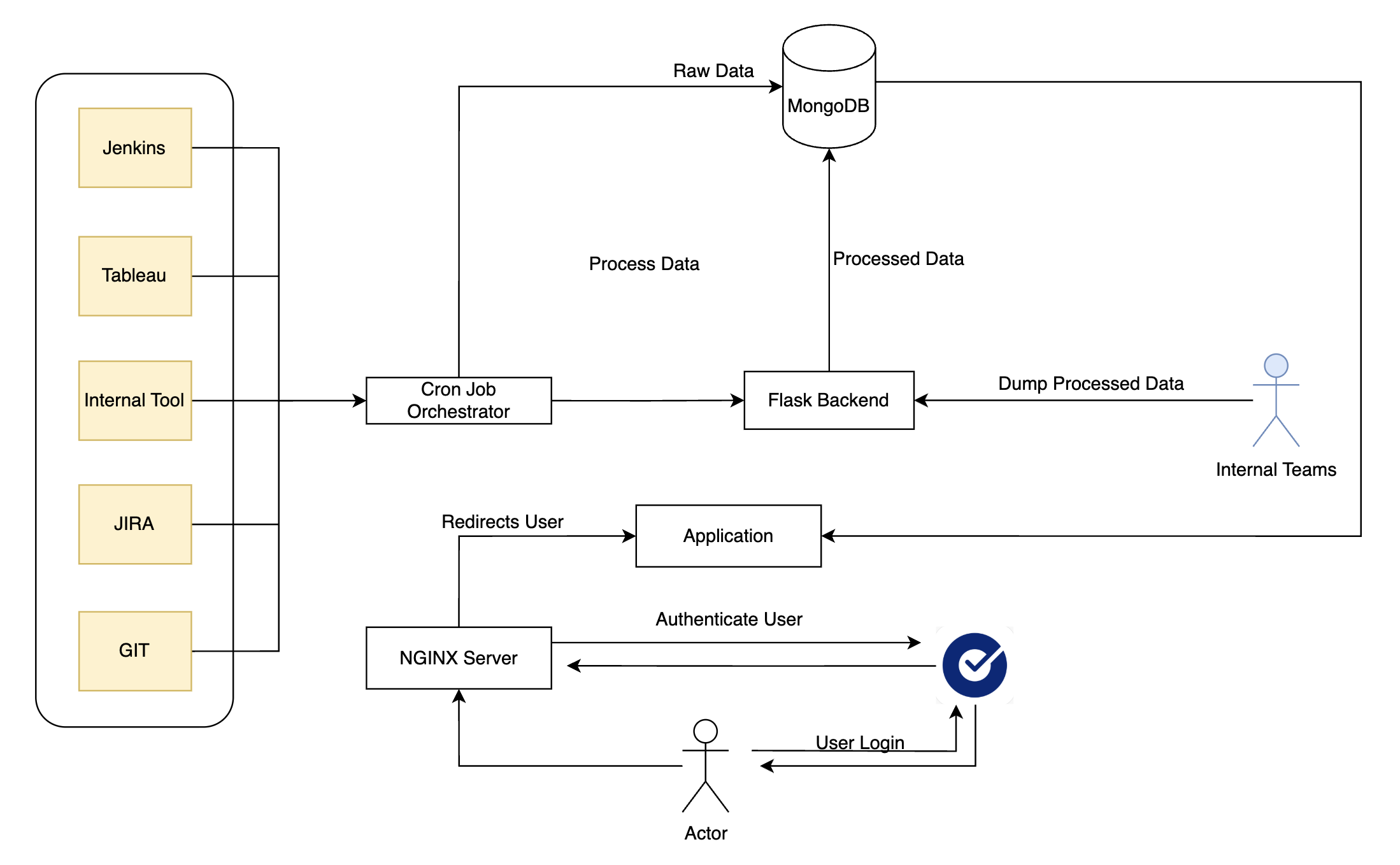}
\caption{High-Level System Architecture. The diagram illustrates the flow of data from various sources through the ingestion, storage, and processing layers to the Application presentation layer, with security and alerting as integral components.}
\label{fig:architecture}
\end{figure*}

The architecture is a modular, multi-layered system designed for scalability and maintainability. It comprises six core layers: Data Ingestion, Data Storage, Data Processing, Presentation, a secure API endpoint, and a cross-cutting Security layer, as illustrated in Fig.~\ref{fig:architecture}.

\subsection{Data Ingestion Layer}
The foundation is a robust and resilient data ingestion layer responsible for fetching raw data from a wide array of sources using a series of cron jobs orchestrated by Jenkins. Each data source has a dedicated job tailored to its volatility. The granular implementation details of the data sources and the system’s error handling mechanisms are described in Appendix A.

\subsection{Data Storage Layer}
All data is stored in a central MongoDB database, chosen for its flexible, document-oriented data model suitable for semi-structured data from API responses and log files \cite{b18}. We employ a dual-schema strategy:
\begin{itemize}
    \item \textbf{Raw Data Collections:} Store unprocessed data, providing a complete historical record for auditing and reprocessing \cite{b10}.
    \item \textbf{Processed Data Collections:} Store computed, time-series data for each metric, optimized for efficient visualization and trend analysis.
\end{itemize}

\subsection{Data Processing Layer}
This layer consists of scripts that run in near-parallel to the ingestion jobs. Its primary responsibilities are to:
\begin{enumerate}
    \item \textbf{Transform and Clean:} Convert raw data into a structured format suitable for analysis (e.g., parsing Git logs, standardizing user IDs).
    \item \textbf{Compute Metrics:} Pre-calculate the key metrics defined in Section~IV. This pre-computation is critical for application performance, acting as a caching layer.
    \item \textbf{Update Timestamps:} Record a "last updated" timestamp for each metric, providing users with clear data freshness visibility.
\end{enumerate}

\subsection{Alerting Layer}
To enable proactive monitoring, an alert layer continuously checks computed metrics against predefined thresholds. As seen in the mockups (Fig.~\ref{fig:ps-android-full}), metrics like 'Crash Rate' have clear target ranges. If a metric breaches its threshold (e.g., 'Visitor Crash Rate' exceeds 5\%), the alert layer automatically generates and sends a notification to designated Slack channels or email distribution lists. This allows teams to respond to issues like a sudden drop in quality or a pipeline failure promptly.

\subsection{Secure Ingestion API}
To enhance the platform's extensibility, a secure API endpoint (authenticated via tokens) is exposed for internal teams. This API allows other services to dump data in a predefined, structured format directly into specific MongoDB collections. By doing so, other teams are empowered to generate their own visualizations in Metabase without needing to build separate data ingestion and processing pipelines, while ensuring data sanctity, structure, and controlled access.

\subsection{Presentation Layer (Metabase)}
For data visualization and exploration, we selected Metabase, an open-source BI tool. The decision was made after evaluating several alternatives, a comparison of which is shown in Table~\ref{tab:bi-tools}. Metabase offers a seamless user interface, direct connectivity to a wide range of data sources including MongoDB, and native support for writing queries. \cite{b5} Its robust support for role-based access control and integration with our single sign-on (SSO) provider, Okta, were critical factors. A key benefit is that it empowers both technical and non-technical users to explore data and create visualizations without extensive training.

\begin{table}[htbp]
\caption{Comparison of BI Tools}
\begin{center}
\begin{tabularx}{\columnwidth}{lXXX}
\toprule
\textbf{Feature} & \textbf{Metabase} & \textbf{Grafana} & \textbf{Tableau Public} \\
\midrule
Cost & Open-Source & Open-Source & Free (Public) \\
Ease of Use & High & Medium & High \\
MongoDB Connector & Native & Plugin & Limited \\
RBAC & Advanced & Advanced & N/A \\
Self-Hosting & Yes & Yes & No \\
SQL Support & Native GUI \& SQL & Requires specific data source support & Yes \\
\bottomrule
\end{tabularx}
\label{tab:bi-tools}
\end{center}
\end{table}

\subsection{Security Layer: Role-Based Access Control (RBAC)}
Security and data privacy are paramount. We implemented a custom authentication layer that integrates with existing SSO infrastructure. This layer manages user roles and permissions, ensuring that sensitive metrics are only visible to authorized personnel. The authentication process leverages SAML, and once authenticated, Metabase generates a session cookie that enforces the user's permissions.

\section{Key Metrics for Engineering Excellence}
The selection of metrics was a deliberate process, combining industry-standard frameworks like DORA with internal goals tailored to the organization's specific needs. To establish these internal KPIs, we conducted a series of structured workshops with engineering leads and product managers from each platform, followed by a survey to validate the most impactful measures of developer friction and product quality. This ensured the final metric set was aligned with both strategic objectives and the day-to-day realities of our development teams. The metrics provide a balanced view across different facets of engineering, ensuring that speed is not pursued at the expense of quality. The metrics are grouped into five distinct "Developer Areas.”

\subsection{Developer Productivity}
These metrics focus on the efficiency and velocity of the development process, from the initial act of writing code to its final integration and deployment. They are designed to identify friction points and measure the direct impact of tooling and process improvements on the developer workflow.
\begin{itemize}
    \item \textbf{Lead Time for Changes:} A core DORA metric, this measures the median time elapsed from a code commit to that code being successfully deployed into production. This is a holistic indicator of the entire delivery pipeline's efficiency, encompassing not just coding and review but also testing, integration, and deployment processes. A shorter lead time signifies a highly automated and efficient path to delivering value. Data is sourced by correlating commit timestamps from Git with deployment timestamps from Jenkins.
    \item \textbf{PR Cycle Time:} Defined as the median time taken from when a Pull Request (PR) is created to when it is merged into the main branch. It is a key indicator of development velocity and highlights potential bottlenecks in the code review and local validation processes. A consistently low PR Cycle Time suggests a healthy and efficient code review culture. Data is sourced from Git.
    \item \textbf{Code Commit Frequency:} This tracks the average number of commits per developer per day to the main development branch. A higher frequency often indicates smaller, more manageable batch sizes, which are a hallmark of continuous integration. It reflects a development team's rhythm and flow, with frequent commits reducing the risk of large, complex merges. Data is sourced directly from Git history.
    \item \textbf{Build Induced Latency:} This metric quantifies the "wait time" developers experience by measuring the median time taken for a continuous integration build to provide feedback after a PR is submitted or updated. This is a direct measure of friction in the inner development loop; long build times interrupt a developer's flow, increase context switching, and slow down iteration. Data is sourced from Jenkins and Splunk.
    \item \textbf{PR Throughput:} The average number of Pull Requests merged per team per week. This metric helps in understanding team capacity and output over time. When analyzed as a trend, it can reveal the impact of process changes or highlight teams that may be under-resourced. Data is sourced from Git.
    \item \textbf{Co-pilot Assistance Metrics:} This modern metric suite is crucial for understanding the impact of AI-assisted coding tools. \cite{b13} It captures metrics such as the number of suggestions provided, acceptance and decline rates, and overall lines of code generated. A high acceptance rate can be an indicator of improved developer productivity and efficiency.
\end{itemize}

\subsection{Quality and Reliability}
This area measures the stability and quality of the product experienced by the end-users.
\begin{itemize}
    \item \textbf{Stability / User Crash Rate:} The percentage of users who have \textit{not} experienced a crash (Stability) or the percentage of sessions that end in a crash (User Crash Rate). This is a critical user-facing metric. Monitoring this helps in quickly identifying releases with regressions that negatively impact user experience. Data is collected from mobile services for native applications and from analytics tools for web applications.
    \item \textbf{Blocker/Critical Bugs:} The number of open bugs with a priority of "Blocker" or "Critical" in Jira. This provides a direct, unfiltered view of high-priority issues that may be preventing users from completing key workflows or causing significant user frustration.
    \item \textbf{Bug Mix Overall:} The distribution of bugs across different severity levels (Blocker, Critical, Major, Minor, Normal). As shown in the weekly and monthly graphs (see Fig.~\ref{fig:ps-android-full}), a healthy trend is a downward slope for high-severity bugs and a stable or declining number of lower-severity bugs.
\end{itemize}

\subsection{Operational Efficiency (DORA Alignment)}
These metrics are aligned with DORA principles and measure the efficiency and reliability of the delivery pipeline.
\begin{itemize}
    \item \textbf{Deployment Frequency:} The number of deployments to production per week. A core DORA metric that indicates the team's ability to deliver value to customers frequently. \cite{b7} A high frequency is indicative of a mature and automated CI/CD pipeline. Data is sourced from an internal deployment tool.
    \item \textbf{Main Fail Rate:} The percentage of builds that fail on the primary (main) branch over a given period. This DORA-inspired metric reflects the stability of the CI/CD pipeline and is sourced from Splunk. \cite{b11}
    \item \textbf{Average Turnaround Time:} The time taken from triggering a build to its successful completion, indicating the efficiency of the build and test process. Data is sourced from Splunk.
\end{itemize}

\subsection{Automation Tooling}
This area focuses on the health, effectiveness, and efficiency of our automated testing infrastructure. The goal of these metrics is to ensure that the automation suite is not just a collection of tests, but a reliable, fast, and trustworthy system that enables developers to release code with high confidence and minimal friction. A healthy automation ecosystem is a direct enabler of high-frequency, low-risk deployments.
\begin{itemize}
\item \textbf{Automation Code Coverage:} The percentage of code statements covered by automated tests (primarily unit and integration tests). While not a direct measure of test quality, it serves as a crucial guardrail to prevent untested code from entering the main branch. Tracking this trend over time helps ensure that test debt is not accumulating. Data is sourced from code coverage tools integrated into the CI/CD pipeline.

\item \textbf{Automation Health:} The percentage of automated test suites that passed in a given build or release cycle. This is a high-level indicator of the entire test suite's reliability. A consistently high health score (e.g., 95\%) builds developer trust in the CI/CD pipeline's feedback.

\item \textbf{Test Automation Status:} This provides a clear view of the automation backlog and progress. It breaks down the total number of testable issues (e.g., from Jira) into distinct categories: \textit{To be automated}, \textit{Automated}, and \textit{Can't be automated}. This metric is crucial for program management to understand automation velocity and resource allocation.
\end{itemize}

\subsection{Engagement}
Engagement metrics help understand how users are interacting with the product.
\begin{itemize}
    \item \textbf{MAU (Monthly Active Users):} The number of unique users who actively engage with the product in a month.
    \item \textbf{Rolling MAU \% (RMAU):} A trended view of user engagement over time, smoothing out short-term fluctuations to understand long-term growth or decline in the active user base. Data for both is sourced from Tableau.
\end{itemize}

\section{Evaluation and Results}
The application introduction has had a significant and measurable impact. We evaluate its success based on quantitative gains, qualitative improvements, and stakeholder feedback.

\subsection{Quantitative Impact}
The most direct quantitative benefit of the framework was the significant reduction in manual effort required for performance reporting across engineering, program management, and leadership. The system automated the time-consuming tasks of data gathering from disparate sources, curating weekly and monthly reports, and, most importantly, identifying cross-platform linkages that were previously obscured.

\begin{itemize}
    \item \textbf{Time Savings:} We estimate a saving of \textbf{40 person-hours per week} across the organization. This figure is primarily based on surveys with 15 engineering and program managers who no longer need to perform manual data collection and reporting. This automation has a cascading benefit: engineers face fewer interruptions for ad-hoc data requests, and leadership can access near-real-time insights directly, eliminating the latency of manual report generation cycles.
\end{itemize}

\subsection{Qualitative Impact: Case Studies}
The framework's true power lies in its ability to correlate data from different sources, revealing previously obscured insights.

\begin{figure}[htbp]
\centering
\begin{tikzpicture}
\begin{axis}[
    width=0.8*\columnwidth,
    height=2.2in,
    xlabel={Date},
    ylabel={PR Cycle Time (Hours)},
    ymode=log,
    yticklabel style={/pgf/number format/fixed},
    legend pos=north west,
    axis y line*=left,
    ymajorgrids=true,
    xmin=1, xmax=14,
    xtick={1,3,5,7,9,11,13},
    xticklabels={Day 1,Day 3,Day 5,Day 7,Day 9,Day 11,Day 13},
    xticklabel style={rotate=45, anchor=east},
]
\addplot[color=blue, mark=*, line width=1.5pt] coordinates {
    (1, 28) (2, 26) (3, 29) (4, 31) (5, 45) (6, 52) (7, 55) (8, 48) (9, 32) (10, 29) (11, 28) (12, 27) (13, 28) (14, 26)
};
\legend{PR Cycle Time}
\node[draw, fill=white, text width=2.5cm, align=center] at (axis cs:9,100) {Test Quarantined};
\draw[->, thick] (axis cs:9,80) -- (axis cs:8.5,50);
\end{axis}
\begin{axis}[
    width=0.8*\columnwidth,
    height=2.2in,
    ylabel={Main Fail Rate (\%)},
    axis y line*=right,
    axis x line=none,
    legend pos=north east,
    ymin=0, ymax=25,
    xmin=1, xmax=14,
]
\addplot[color=red, mark=square*, dashed, line width=1.5pt] coordinates {
    (1, 4) (2, 5) (3, 4) (4, 6) (5, 15) (6, 20) (7, 18) (8, 16) (9, 7) (10, 5) (11, 4) (12, 5) (13, 4) (14, 4)
};
\legend{Fail Rate}
\end{axis}
\end{tikzpicture}
\caption{Correlation of PR Cycle Time and Main Fail Rate during a CI/CD incident, illustrating the application's diagnostic power.}
\label{fig:case-study-chart}
\end{figure}
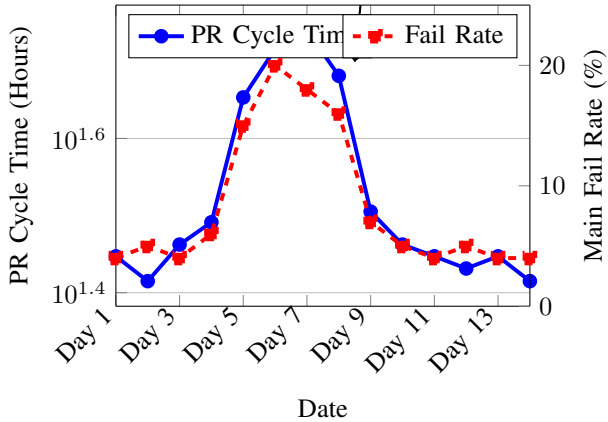

\textbf{Bottleneck Identification:} In one instance, the application highlighted a sudden 30\% increase in 'PR Cycle Time' for a specific platform. By cross-referencing this with other metrics, managers noticed a simultaneous spike in the 'Main Fail Rate'. This correlation, visualized in Fig.~\ref{fig:case-study-chart}, immediately suggested that CI pipeline failures were blocking PR merges. A drill-down into Splunk data revealed a flaky integration test. The test was quarantined, and 'PR Cycle Time' returned to normal within \textbf{48 hours}. Without the unified view, this root cause analysis could have taken days.

\textbf{Productivity Tool Impact Analysis:} After integrating GitHub Copilot, the application showed a clear uptick trend in 'PR Throughput' for early adopter teams, providing a strong data-driven justification for expanding its use across the organization.

\subsection{Stakeholder Feedback}
The reception has been overwhelmingly positive. A Senior Engineering Manager noted, \textit{"For the first time, we have a single pane of glass to understand the health of our engineering organization. The ability to go from a high-level trend to a specific failing build in two clicks is a game changer."}

\subsection{Application Visualizations}
\begin{figure*}[htbp]
    \centering
    \includegraphics[width=\textwidth]{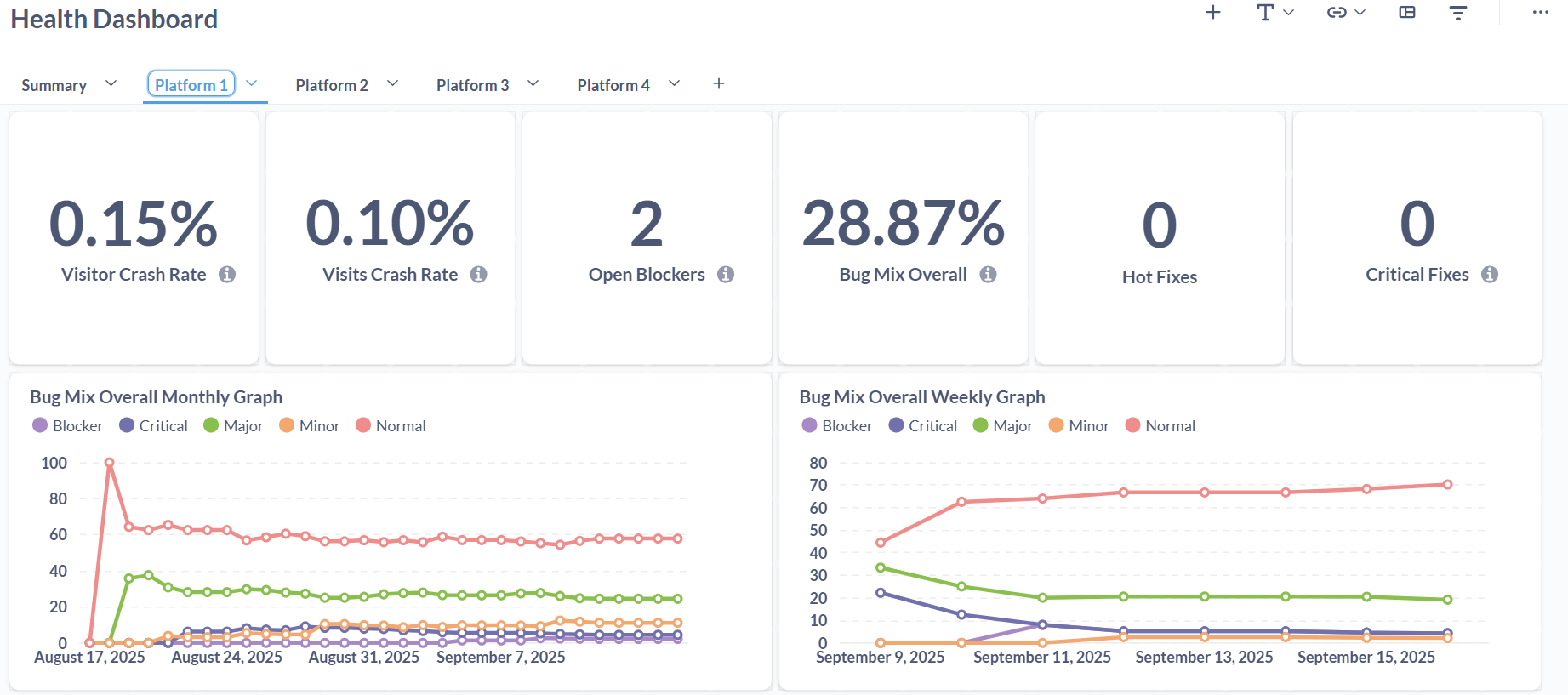}
    \caption{Application View for a Digital Platform. This view consolidates key quality and reliability metrics, including Blockers, Fixes, Crash Rate, and the overall Bug Mix, providing an at-a-glance health assessment.}
    \label{fig:ps-android-full}
\end{figure*}

\begin{figure*}[htbp]
    \centering
    \includegraphics[width=\textwidth]{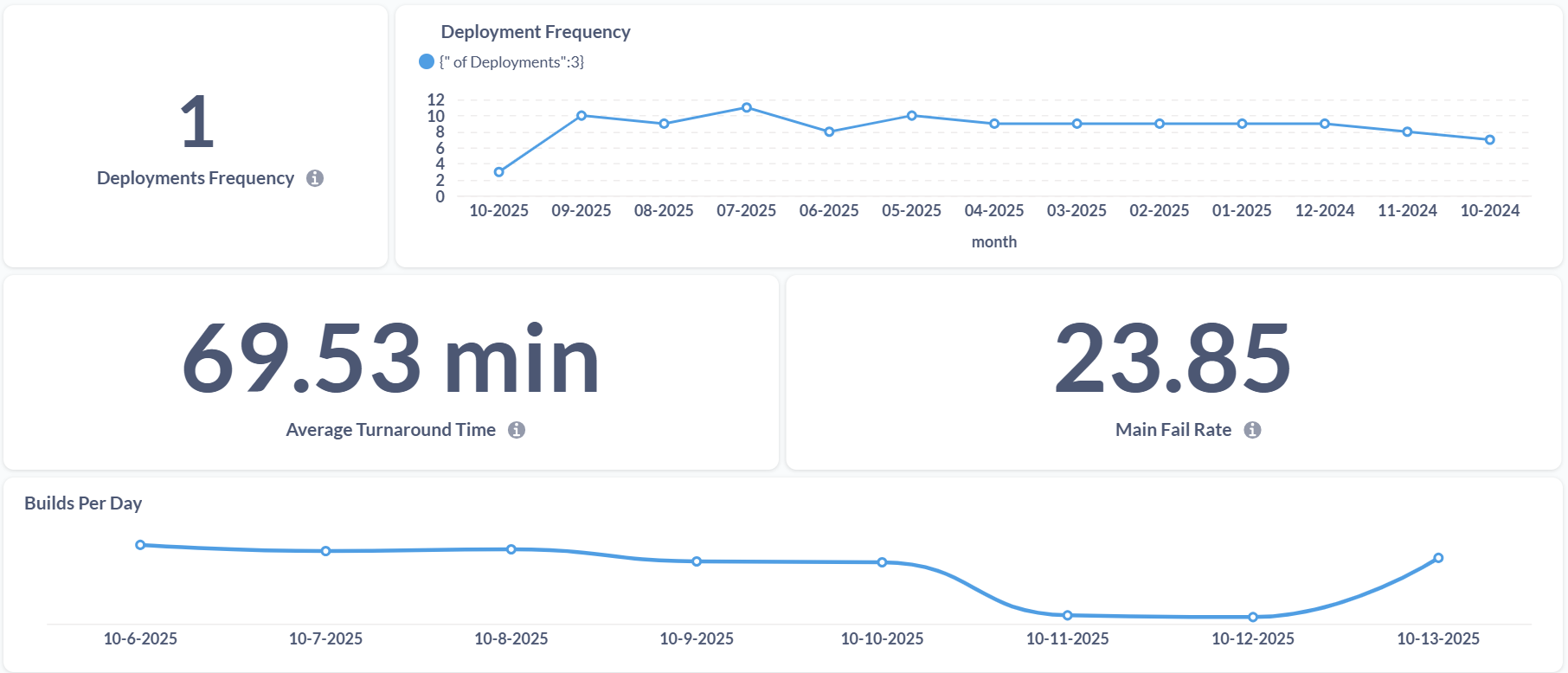}
    \caption{Application View for a Digital Platform. This visualization focuses on operational efficienct metrics such as Deployment Frequency and Fail Rate.}
    \label{fig:ps-web-full}
\end{figure*}

The figures below show example visualizations from the application. Fig.~\ref{fig:ps-android-full} shows a dashboard consolidating key quality and reliability metrics, while Fig.~\ref{fig:ps-web-full} focuses on operational efficiency metrics. The combination of single value metrics for at-a-glance health checks and time-series graphs provides a comprehensive view.

\section{Discussion}
The implementation surfaced important considerations regarding scalability, performance, platform governance, and the responsible use of metrics.

\subsection{Scalability}
The current architecture is designed to be scalable. Onboarding a new metric or data source is a streamlined process. However, we have identified two potential scalability bottlenecks: the cron job orchestrator and the DB instance. As data volume grows, running an excessive number of serial cron jobs could lead to delays. Similarly, the database could become bloated if historical data is not managed properly. To mitigate this, the architecture can be scaled by parallelizing cron jobs. Metabase's ability to connect to multiple data sources simultaneously also means we can federate queries across different databases if needed, reducing the load on a single instance.

\subsection{Performance Optimization: The Role of Caching}
Application's latency is a critical factor for user experience. Our architecture emphasizes pre-computation. The data processing layer acts as a caching mechanism, calculating complex metrics and storing the final values in MongoDB. Metabase then simply queries this pre-computed value, resulting in near-instantaneous load times. Furthermore, Metabase's open-source edition provides a built-in caching feature. We leverage this to cache the results of frequently run queries for a specified time-to-live (TTL).

\subsection{Threats to Validity}
We acknowledge several potential threats to the validity of our evaluation.
\begin{itemize}
    \item \textbf{Internal Validity:} Our case studies are observational and do not represent controlled experiments. While the correlation between the CI failure and the increase in `PR Cycle Time` is strong, other unobserved factors could have contributed to this effect.
    \item \textbf{External Validity:} The framework was designed within the context of a large, multi-platform software organization. The specific metric choices, data sources, and architectural decisions may not be directly generalizable to smaller organizations or different software domains (e.g., embedded systems vs. web services) without adaptation.
    \item \textbf{Construct Validity:} We measure abstract concepts like "productivity" using proxy metrics (e.g., `PR Throughput`). These proxies are imperfect and can be influenced by external factors. This is closely related to the risk of Goodhart's Law ("when a measure becomes a target, it ceases to be a good measure"), which we actively mitigate by focusing on a balanced set of metrics rather than optimizing for any single one.
\end{itemize}

\subsection{Ethical Considerations and Metric Misuse}
Implementing any system that measures developer activity requires careful consideration of its cultural impact. Our guiding principle was that this framework must be a tool for empowerment and process improvement, not for individual performance evaluation. We mitigate the risk of metric misuse in several ways:
\begin{enumerate}
    \item \textbf{Team-Level Aggregation:} Metrics related to individual output, such as 'Code Commit Frequency,' are always aggregated at the team or organization level, never displayed for individuals.
    \item \textbf{Leadership Training:} We conducted sessions with engineering managers to frame the metrics as diagnostic tools for identifying systemic bottlenecks, explicitly discouraging their use in performance reviews.
    \item \textbf{Focus on Trends over Absolutes:} We encourage stakeholders to focus on trends and relative changes for a team over time, rather than comparing absolute values between teams with different contexts and charters.
\end{enumerate}

\subsection{Limitations of Metabase}
While Metabase is a powerful open-source tool, its visualization options are not as extensive as some commercial BI tools. Performance can also degrade when non-technical users attempt ad-hoc queries on very large raw datasets, sometimes leading to query timeouts or browser performance issues. \cite{b23}

\section{Future Work}
Our future research agenda is focused on enhancing the intelligence of the platform.
\begin{itemize}
    \item \textbf{Anomaly Detection:} We plan to move beyond static, threshold-based alerting by implementing machine learning models (e.g., ARIMA or LSTM networks) to automatically detect statistically significant anomalies in key metrics. This builds on established research in time-series anomaly detection \cite{b_chandola_2009}.
    \item \textbf{Predictive Analytics:} We will leverage historical data to build predictive models that can forecast project delays or identify teams at risk of burnout based on trends in PR throughput and cycle time. The validity of these models will be tested against qualitative data from team surveys.
    \item \textbf{Deeper Integration:} We aim to create more interactive UI that allow users to take action directly from the view, such as clicking a "stale PR" metric to trigger a Slack reminder to the assigned reviewers, closing the loop between insight and action.
\end{itemize}

\section{Conclusion}
The centralized engineering performance framework has successfully transformed how the organization monitors and improves its software development lifecycle. By consolidating data from disparate sources into a single, cohesive view, the platform has eliminated hours of manual report generation, enabled faster, data-driven decision-making, and fostered a culture of shared ownership over key performance indicators. The architecture, built on a foundation of resilient data ingestion, pre-computation, and a flexible open-source BI tool, has proven to be both scalable and performant. The thoughtful selection of DORA-aligned and internally defined metrics provides a comprehensive understanding of developer productivity, product quality, and operational efficiency. This work demonstrates that a well-architected analytics platform is an indispensable tool for any large-scale engineering organization aiming for continuous improvement and engineering excellence.

\appendix
\section{Implementation Details of the Data Ingestion Layer}
\subsection{Data Sources and Fetching Schedules}
\begin{itemize}
    \item \textbf{Git:} Daily cron jobs extracts commit history, branch information, and pull request metadata.
    \item \textbf{Jira:} Fetched daily to gather data on issue types (bugs, stories), priorities, statuses, and resolution times.
    \item \textbf{Splunk:} High-frequency jobs (e.g., hourly) query Splunk for application logs, build failures, and performance data.
    \item \textbf{Crash reporting service:} Fetched daily for detailed, symbolicated crash reports and stability metrics.
    \item \textbf{Mobile app store console:} Daily jobs connect via API to fetch Android-specific quality metrics like ANR rates.
    \item \textbf{Tableau:} Daily jobs pull pre-aggregated business metrics, such as Monthly Active Users (MAU).
    \item \textbf{Other Sources:} Data from internal deployment tools, web analytics, and custom APIs are pulled at corresponding frequencies.
\end{itemize}

\subsection{Resiliency and Error Handling}
The ingestion layer is designed for fault tolerance. If a Jenkins cron job fails, an automated retry mechanism is triggered up to five times with increasing backoff intervals. Should the failure persist, an alert is generated on a dedicated developer Slack channel, and the system uses the last successfully fetched data to prevent application downtime. To handle API rate limits, we respect `Retry-After` headers and distribute requests evenly. The system also monitors credential expiration and sends proactive alerts to administrators to ensure uninterrupted data flow.

\end{document}